\documentclass[prb,aps]{revtex4}

\usepackage{graphicx}
\usepackage{amssymb}
\usepackage{amsthm}
\usepackage{amsmath}
\usepackage{bbold}
\usepackage{color}
\newcommand{\av}[1]{\left\langle #1 \right\rangle} 
\DeclareMathOperator{\trace}{Tr}

\DeclareMathOperator{\sign}{sign}
\DeclareMathOperator{\Log}{Log}
\def\TrLog{\trace\Log}
\newcommand{\mc}{\mathcal}

\begin{document}

\title{Using Majorana spin-1/2 representation for the spin-boson model}
\author{Pablo Schad}
\affiliation{Institut f\"ur Theorie der Kondensierten Materie,
Karlsruhe Institute of Technology, 76128 Karlsruhe, Germany}
\author{Alexander Shnirman}
\affiliation{Institut f\"ur Theorie der Kondensierten Materie,
Karlsruhe Institute of Technology, 76128 Karlsruhe, Germany}
\author{Yuriy Makhlin}
\affiliation{L.D. Landau Institute for Theoretical Physics, acad. Semyonov 
av., 1a, 142432, Chernogolovka, Russia}
\affiliation{Moscow Institute of Physics and Technology, 141700, Dolgoprudny, Russia}

\begin{abstract}
In this paper we resolve a contradiction between the fact that 
the method based on the Majorana representation of spin-1/2 is exact and its
failure to reproduce the perturbative Bloch-Redfield relaxation rates.  
Namely, for the spin-boson model, direct application of this method in the leading order allows for
a straightforward
computation of the transverse-spin correlations, however, for the longitudinal-spin correlations it
apparently fails in the long-time limit. Here we indicate the reason for this failure. 
Moreover, we suggest how to apply this method to allow, nevertheless, for simple and accurate 
computations of spin correlations. Specifically, we
demonstrate that accurate results are obtained by avoiding the use of the longitudinal Majorana
fermion, and that correlations of the remaining transverse Majorana fermions can be easily evaluated
using an effective Gaussian action.

\end{abstract}

\maketitle

\section{Introduction}
\label{sec:intro}

Application of field-theoretical methods to spin systems is hampered by the non-Abelian
nature of the spin operators~\cite{altland}. To circumvent this problem, one can represent spin
operators in terms of fermions or bosons and use standard field theory~\cite{Tsvelik}.
Several formulations have been suggested, including the Jordan-Wigner~\cite{JordanWigner} and
Holstein-Primakoff~\cite{Holstein} transformations, the Martin~\cite{Martin} Majorana-fermion and
Abrikosov~\cite{abrikosov} fermion representations as well as the
Schwinger-boson~\cite{schwinger,arovas,read,wang} and
slave-fermion~\cite{affleck,marston,andrei,dagotto,wen,nayak} techniques.

Among other representations of spin operators the Majorana-fermion approach has a special property.
In most other approaches the necessary extension of the Hilbert space requires taking into account
additional constraints on the boson/fermion operators, which place the system to the physical
subspace. As it was realized early on~\cite{Spencer68} and reemphasized recently~\cite{Schad2015},
for the Majorana representation this complicating step is not needed. 
Moreover, it was further observed~\cite{Mao03,Shnirman03} that a wide class of spin
correlation functions can be reduced to Majorana correlations of the same order, so that the spin
correlation functions can be computed very efficiently. This, in particular, includes
single-spin problems, like the Kondo problem or the spin-boson problem.
In this case one avoids potentially complex vertex corrections to the external vertices, and
spin-spin correlation functions are essentially given by single-particle fermionic lines rather
than by loop diagrams.

Despite these advantages, the Majorana spin-1/2 representation is not used very often. One possible
reason is that its application requires careful calculations. In particular, for the spin-boson
model, application of this method allows for a straightforward computation of the correlation
functions of the transverse spin components~\cite{Shnirman03}, using the lowest-order self-energy.
However, for longitudinal-spin correlations, it fails in the long-time limit. Here we indicate the
reason for this behavior. Moreover, we suggest an approach that allows one to use the
Majorana-fermion representation as a
convenient and accurate tool for generic spin correlations. More specifically, we demonstrate that
accurate results are obtained in low orders by avoiding the use of the longitudinal Majorana fermion
operator. Furthermore, we demonstrate that correlations of the remaining, transverse Majorana
fermions can be easily evaluated using an effective Gaussian action.

Here we show, how this problem arises and how it can be solved. First, we do so, using the standard
diagrammatic methods. Then we reformulate the problem, using the path-integral formalism. This
path-integral analysis allows us also to formulate a generalized Wick theorem and to provide a
prescription for efficient computation of spin correlation functions. 
Finally, we demonstrate that the method indeed produces the expected longitudinal-spin correlations.


\section{Spin-Boson Model}
\label{sec:model}

The spin-boson model~\cite{LeggettSB87} describes a spin-1/2 in a finite magnetic field. One of the transverse components of the spin is coupled to the bosonic bath. The Hamiltonian  reads
\begin{equation}
\label{SBModel}
H = B \hat{S}^z +  \hat X\hat{S}^x + H_B\,,
\end{equation}
where $H_B$ is the Hamiltonian of the bosonic bath, which governs the free dynamics of $\hat X$. 
The bath is characterized by the correlation function $\Pi(\tau-\tau')\equiv \langle T X(\tau)X(\tau')\rangle$. 
In the Matsubara representation the correlation function $\Pi(\tau-\tau')$ can be written as 
\begin{equation}
\Pi(i\omega_m) = \int\limits_{-\Lambda}^{\Lambda} \frac{d x}{\pi} \,\frac{\rho(|x|) \sign x}{x - i\omega_m} \ ,
\end{equation}
where $\rho(|x|)$ is the bath spectral density, and $\omega_m = 2\pi m T$. In the Ohmic case
considered here $\rho(|x|) = g |x|$. For $\omega_m\ll \Lambda$ this gives $\Pi(i\omega_m) \approx
(2g/\pi)\Lambda -g |\omega_m|$. 

In the weak coupling limit, $g\ll 1$, the spin-boson model
can be solved very efficiently by the master equation (Bloch-Redfield) technique~\cite{Bloch57,Redfield57}. In this limit the
dynamics is characterized by two rates, $\Gamma_1$ and $\Gamma_2 = \Gamma_1/2$. The dephasing rate
$\Gamma_2$ describes the relaxation of the transverse spin components $\hat S_x$ and $\hat S_y$ to
their equilibrium values, $\langle S_x \rangle =\langle S_y \rangle=0$. The rate $\Gamma_1$
describes the relaxation of the longitudinal component $\hat S_z$ to its equilibrium value  $\langle
S_z \rangle = (1/2) \tanh[\beta B/2]$. One readily obtains 
\begin{equation}
\label{eq:Gamma12}
\Gamma_1=2\Gamma_2 = \frac{gB}{2}\coth[\beta B/2]\ .
\end{equation}
From this one immediately concludes that the symmetrized, real-time correlation functions of the
transverse spin components, $2\langle \left\{S_x(t_1),S_x(t_2)\right\}_+\rangle$ and $2\langle
\left\{S_y(t_1),S_y(t_2)\right\}_+\rangle$, are given in the frequency domain by Lorentzians of unit
weight and width $\Gamma_2$. In contrast, the longitudinal correlation function $2\langle
\left\{S_z(t_1),S_z(t_2)\right\}_+\rangle$ consists in the Fourier representation 
of a delta function (zero-width Lorentzian) of weight $4 \langle S_z \rangle^2$ and a Lorentzian of width $\Gamma_1$ and 
weight $1- 4 \langle S_z \rangle^2$. In Ref.~\onlinecite{Shnirman03} an attempt was made to employ the 
Majorana representation described in Appendix~\ref{sec:majrep} and, specifically, the reduction (\ref{eq:Trick})
to reproduce these well established facts. For the transverse correlation functions these
results were indeed reproduced, and the calculation was significantly simpler than any of those
attempting to calculate the fermionic loop. Yet, for the longitudinal 
correlations a result was obtained, which failed in the long-time limit. Namely, instead of
two Lorentzians (one with zero width and the other with width $\Gamma_1$) a single Lorentzian with
the width $\tilde \Gamma_1 = \Gamma_1 \left(1-\tanh^2[\beta B/2]\right)$ and unit weight was
found. This solution reproduces correctly the short-time ($|t_1-t_2| \ll 1/\Gamma_1$) 
decay of the longitudinal correlation function, but fails in the long-time limit. Indeed, the wrong
solution decays to zero, whereas the correct one should decay to $4 \langle S_z \rangle^2$.

The reason for the failure is intuitively clear. In Eq.~(\ref{Majrep2}) the spin operator $\hat S_z$ is essentially represented 
by the Majorana operator $\hat \eta_z$. The latter, however, is zero on average. Yet,
since the Majorana representation is definitely valid, a resolution of the problem is needed. 

\section{Diagrammatic analysis}
\label{sec:diagrams}

In this section we show that from the technical viewpoint the discussed failure
of the diagrammatic calculation at long times arises from the insufficiency of the lowest-order
approximation for the self-energy. In Section~\ref{sec:Wick} we propose a method to circumvent
this problem. 

In the Majorana representation~\eqref{Majrep} the Hamiltonian of the system reads
\begin{align}
 H= B \hat{S}^z +  \hat{S}^x\,\hat X + H_B = - i B \,\hat\eta_x \hat\eta_y - i \hat\eta_y \hat\eta_z\, \hat X + H_B\ .
\label{Hint}
\end{align}
We employ the standard diagrammatic rules for calculations perturbative in the spin-bath
interaction. The zeroth-order Green function of the Majorana fermions is obtained from 
\begin{align}\label{eq:G0inv}
 [G^{0}]^{-1} &=  \left(
 \begin{array}{ccc}
- \delta(\tau-\tau') \partial_{\tau'} & i B \delta(\tau-\tau') & 0 \\
 -i B \delta(\tau-\tau') & -\delta(\tau-\tau') \partial_{\tau'} & 0 \\
 0 & 0 & -\delta(\tau-\tau') \partial_{\tau'}
 \end{array}
 \right)\ .
 \end{align}
The Green function is given by the Dyson equation $[G]^{-1} =  [G^{0}]^{-1} -
\hat \Sigma$, where the matrix of the self-energies $\hat \Sigma$ has only two diagonal components,
$\Sigma_{yy} \equiv \Sigma_y$ and $\Sigma_{zz} \equiv \Sigma_z$.

\subsection{First order}

\begin{figure}
\includegraphics[scale=0.3]{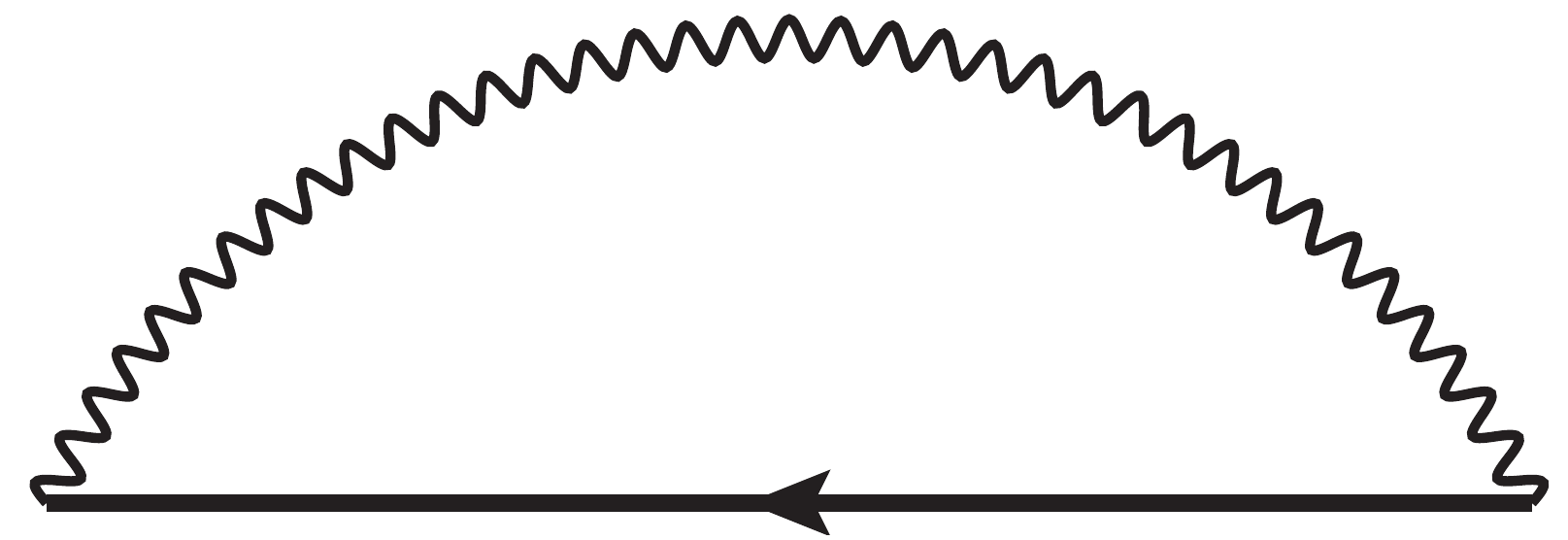}
\caption{The first-order diagram for the self-energy. It also corresponds to the saddle-point self-energy in the path-integral approach. 
Solid line: fermionic Green function, wavy line: bosonic (bath) correlator.}
\label{Sigma1Order}
\end{figure}

We start by calculating the self-energies in the first order in $g$.  Diagrammatically this corresponds to Fig.~\ref{Sigma1Order}. 
We obtain 
\begin{align}\label{Sigma1y}
\Sigma^{(1)}_{y}(\tau-\tau')&=\Pi(\tau-\tau') G^{0}_{zz}(\tau-\tau') \ , 
\end{align}
\begin{align}\label{Sigma1z}
\Sigma^{(1)}_{z}(\tau-\tau')&=\Pi(\tau-\tau') G^{0}_{yy}(\tau-\tau') \ .
\end{align}
We use
\begin{equation}
G^{0}_{zz}(\epsilon_n) = \frac{1}{i\epsilon_n}\quad,\quad G^{0}_{yy}(\epsilon_n) =\frac{1}{2}\left[ \frac{1}{i\epsilon_n-B} +  \frac{1}{i\epsilon_n+B}\right]\ .
\end{equation} 
This gives
\begin{equation}\label{Sigma1yResult}
\Sigma^{(1)}_{y}(i\epsilon_n)  =-\frac{1}{2} \int\limits_{-\Lambda}^{\Lambda} \frac{dx}{\pi}\,\frac{gx\coth[\beta x/2]}{x-i\epsilon_n}\ .
\end{equation}
and
\begin{equation}\label{Sigma1zResult}
\Sigma^{(1)}_{z}(i\epsilon_n)  =-\frac{1}{4}\sum_{\gamma = \pm 1} \int\limits_{-\Lambda}^{\Lambda} \frac{dx}{\pi}\,
\frac{gx\left[\coth[\beta x/2]-\tanh[\beta\gamma B/2]\right]}{x-i\epsilon_n-\gamma B}\ .
\end{equation}

Upon analytic continuation we obtain the retarded self-energies. For the transverse component we
find
\begin{equation}\label{Sigma1yRetarded}
{\rm Im}\Sigma^{(1)}_{y}(i\epsilon_n \rightarrow \pm B + i0)  = -\Gamma_1  \,,
\end{equation}
where $\Gamma_1$ is the standard longitudinal relaxation rate given by Eq.~(\ref{eq:Gamma12}).
Note that upon substitution to the Dyson equation $[G]^{-1} =  [G^{0}]^{-1} - \hat \Sigma$ this
gives the following denominator for the transverse Green function: 
$G_\perp(\epsilon) \sim \left[\epsilon(\epsilon+i\Gamma_1)-B^2\right]^{-1}$. Thus the poles have the imaginary part 
$\approx -\Gamma_1/2 = -\Gamma_2$, corresponding to the standard dephasing rate.

For the longitudinal component we obtain
\begin{equation}
{\rm Im}\Sigma^{(1)}_{z}(i\epsilon_n\rightarrow 0+i0)  = - \tilde \Gamma_1\ ,
\end{equation}
where 
\begin{equation}
\tilde \Gamma_1 = \Gamma_1 \left(1-\tanh^2[\beta B/2]\right) \ .
\end{equation}
Thus we reproduce the problem noticed in Ref.~\onlinecite{Shnirman03}. Instead of the anticipated solution of the type 
\begin{equation}\label{eq:Ganticipated}
G_{zz,anticipated}^R(\epsilon) = \frac{A}{\epsilon+i0} + \frac{1-A}{\epsilon+i\Gamma_1} \ , 
\end{equation}
where $A \equiv (2 \langle S_z \rangle)^2 =  \tanh^2[\beta B/2]$, we obtain 
\begin{equation}\label{eq:Gobtained}
G_{zz,obtained}^R(\epsilon) = \frac{1}{\epsilon+i(1-A)\Gamma_1} \ .
\end{equation}
One can easily observe that these solutions coincide in the limit $\epsilon \gg \Gamma_1$. Thus our
lowest-order evaluation of the self-energy catches correctly the short-time limit, whereas the
long-time limit ($t \gg 1/\Gamma_1$) needs further analysis. 

\subsection{Second order}

We first compare the self-energies corresponding to (\ref{eq:Ganticipated}) and
(\ref{eq:Gobtained}). For the anticipated solution we obtain 
\begin{equation}\label{eq:SigmaCorrect}
\Sigma^{anticipated}_{z}(\epsilon)  = -i \,\frac{(1-A) \Gamma_1}{1+\frac{iA\Gamma_1}{\epsilon}} =- i
(1-A) \Gamma_1  - \frac{A(1-A)\Gamma_1^2}{\epsilon+iA\Gamma_1} \,,
\end{equation}
whereas in the lowest order we find $\Sigma^{(1)}_{z}(\epsilon)  = - i (1-A) \Gamma_1$. We again
observe the correspondence of the two for $|\epsilon| \gg \Gamma_1$.
We expand
(\ref{eq:SigmaCorrect}) in powers of $g$ and obtain 
\begin{equation}\label{eq:SigmaExpanded}
\Sigma^{anticipated}_z(\epsilon)  = - i (1-A) \Gamma_1  - \frac{A(1-A)\Gamma_1^2}{\epsilon} +
\dots\ .
\end{equation}
Note that for $\epsilon < A\Gamma_1$ this series diverges and the expansion parameter 
in Eq.~\eqref{eq:SigmaExpanded} is $A\Gamma_1/\epsilon$.

\begin{figure}
\includegraphics[scale=0.3]{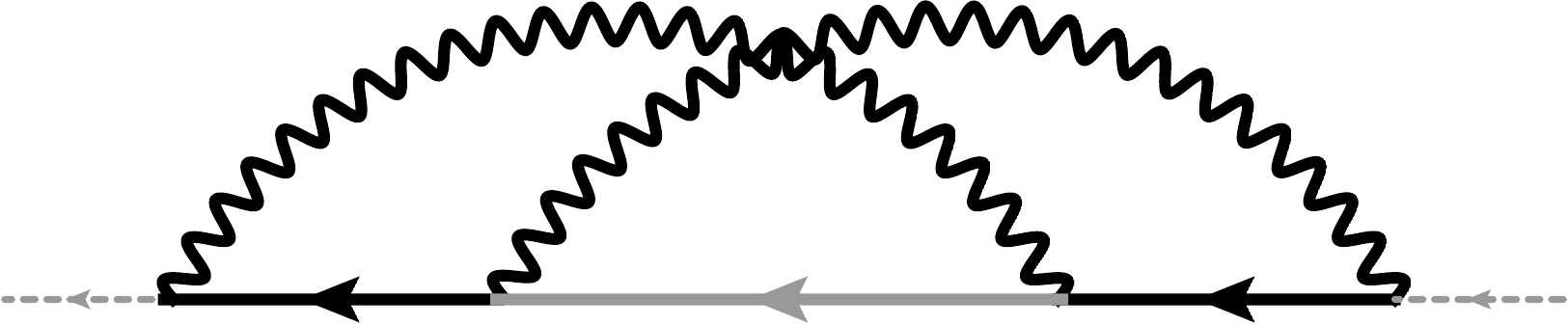}
\includegraphics[scale=0.3]{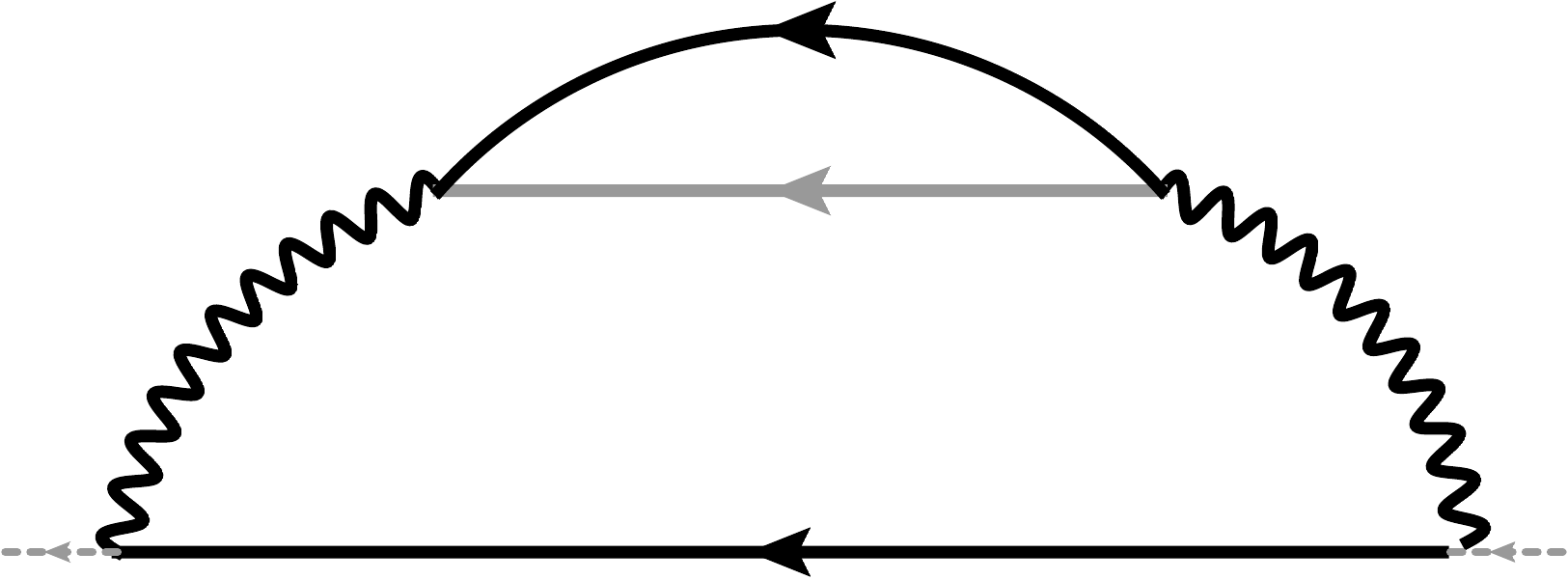}
\includegraphics[scale=0.3]{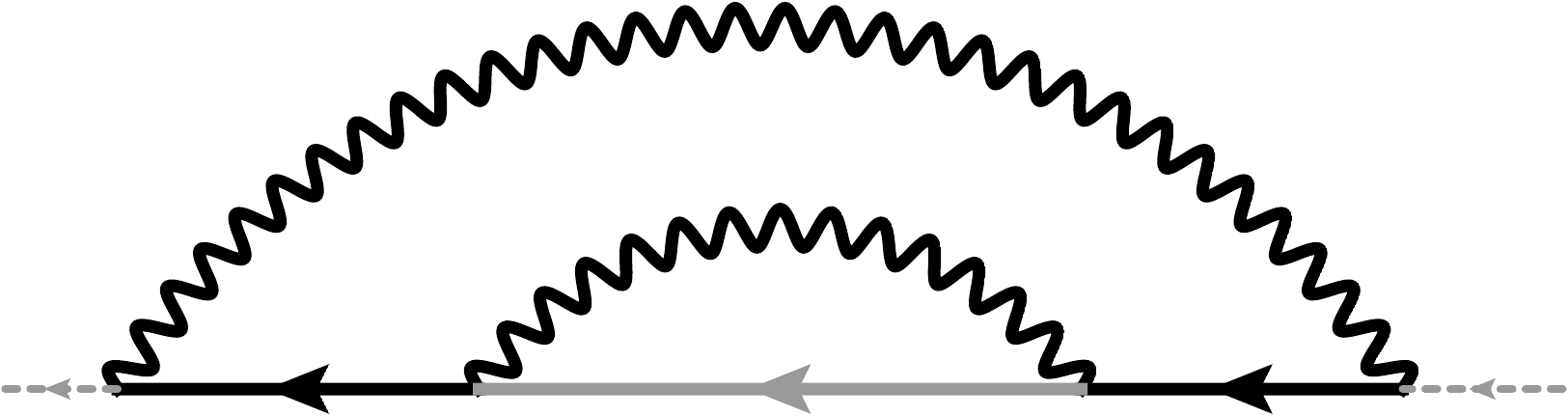}
\caption{Second-order diagrams for self-energy. The colors indicate the two flavors of Majorana
fermions ($y$ and $z$).}
\label{Sigma2Order}
\end{figure}

Eq.~\eqref{eq:SigmaExpanded} suggests that the higher order in $g$ terms in the expansion of the 
self-energy are singular. Thus we attempt to obtain (\ref{eq:SigmaExpanded}) perturbatively in $g$. 
Below we calculate the next, second-order term of this expansion from the relevant
second-order diagrams, which are shown in Fig.~\ref{Sigma2Order}.
For the first (leftmost in Fig.~\ref{Sigma2Order}) diagram we obtain 
\begin{equation}
\Sigma^{(2,1)}_{z}(\epsilon) = \frac{1}{\beta^2}\sum_{\omega_1,\omega_2}\,G_{yy}(\epsilon - \omega_1)G_{zz}(\epsilon-\omega_1 - \omega_2)
G_{yy}(\epsilon-\omega_2) \Pi(\omega_1)\Pi(\omega_2) \ .
\end{equation}
The second (central in Fig.~\ref{Sigma2Order}) diagram is given by 
\begin{equation}
\Sigma^{(2,2)}_{z}(\epsilon) = \frac{1}{\beta^2}\sum_{\omega_1,\omega_2}\,G_{yy}(\epsilon - \omega_1)G_{zz}(\omega_1 + \omega_2-\epsilon)
G_{yy}(\epsilon-\omega_2) \Pi(\omega_1)\Pi(\omega_1) \ .
\end{equation}
Both diagrams strongly diverge ($\propto \Lambda^2$), but their sum diverges at most logarithmically. Indeed it can be written as 
\begin{eqnarray}
\Sigma^{(2,12)}_{z}(\epsilon) &&= \Sigma^{(2,1)}_{z}(\epsilon)+\Sigma^{(2,2)}_{z}(\epsilon)\nonumber\\&&=
-\frac{1}{2\beta^2}\sum_{\omega_1,\omega_2}\,G_{yy}(\epsilon - \omega_1)G_{zz}(\epsilon-\omega_1 - \omega_2)
G_{yy}(\epsilon-\omega_2)\left[\Pi(\omega_1)-\Pi(\omega_2)\right]^2 \ .
\end{eqnarray}
Evaluating this expression (with the help of Mathematica), we reproduced the second term of
(\ref{eq:SigmaExpanded}). All other diagrams, i.e., the third diagram (rightmost in
Fig.~\ref{Sigma2Order}) for $\Sigma_{z}$ and all diagrams for $\Sigma_{y}$ do not lead to
$1/\epsilon$ divergences. 

Thus we reproduce the anticipated expression for the longitudinal self-energy (\ref{eq:SigmaCorrect}) up to the second order in $g$. 
This shows that the discrepancy noticed in Ref.~\onlinecite{Shnirman03} is removed by accounting for higher-order contributions to the self-energy of the longitudinal Majorana fermions. Thus we confirm the importance of higher-order corrections anticipated in Ref.~\onlinecite{Shnirman03} and also discussed in Ref.~\onlinecite{FlorensPRB11}.

For a better understanding of this result, in the next section we develop a path-integral description (see also Refs.~\onlinecite{VieiraPRB81,VieiraPhysA82}). 

\section{Path integral formulation}
\label{sec:path_integral}

We use the Matsubara imaginary-time technique ($t= -i\tau$, $\partial_\tau = -i \partial_t$). 
The partition function, $Z = \int D[\dots] \exp{[iS]}$, reads 
\begin{align}
Z &= \int D[X] D[\eta_\alpha] \, \exp\left\{i \,S_B + \int\limits_0^{1/T} d\tau \left[-\frac{1}{2}\,\eta_\alpha(\tau)\, \partial_\tau \eta_\alpha(\tau) + iB \eta_x \eta_y + i \eta_y \eta_z \,X\right] \right\}\ . 
\end{align}
Here $S_B$ is the free bosonic action. 
The first step is to average over the fluctuations of $X$, yielding 
\begin{eqnarray}
 Z  = &&\int D[\eta_\alpha] \, \exp\left\{\int d\tau \,\left[- \frac 12\,\eta_\alpha(\tau) \,
\partial_{\tau} \, \eta_\alpha(\tau) + i \,B \eta_x(\tau)\eta_y(\tau) \right]\right.
 \nonumber \\
 &&-\left. \frac 12  \int d\tau d\tau' \, \Pi(\tau-\tau') \, \eta_{y}(\tau)
\eta_{z}(\tau) \,  \eta_{y}(\tau') \eta_{z}(\tau') \right\}\ .
\end{eqnarray}
Next we decouple the quartic Majorana-Majorana interaction in a different channel. To this end we
rearrange 
\begin{align}
i S_{int}[\eta_\alpha]&= - \frac 12  \int d\tau d\tau' \, \Pi(\tau-\tau') \, \eta_{y}(\tau) \eta_{z}(\tau) \,  \eta_{y}(\tau') \eta_{z}(\tau') \notag\\
 &= \frac 12 \int d\tau d\tau'  \, \Pi(\tau- \tau') \, \left[ \eta_{y}(\tau) \eta_{y}(\tau')\right] 
 \left[\eta_{z}(\tau) \eta_{z}(\tau')\right] \ .
 \label{Smint}
\end{align}
We now  employ the Hubbard-Stratonovich transformation by introducing the fields $\Sigma_y$ and
$\Sigma_z$. These fields inherit the symmetry of the Majorana propagators, therefore
$\Sigma_\alpha(\tau,\tau')=-\Sigma_\alpha(\tau',\tau)$. The new effective action reads
\begin{eqnarray} \label{SetaQMMM}
 i S[\eta_\alpha,\Sigma_\alpha] =
 \frac{1}{2}\, \int d\tau d\tau'  \,
\eta_\alpha(\tau)\, 
 \left( G^{-1} \right)_{\alpha,\beta}\, \eta_\beta(\tau') 
 - \frac12\, \int d\tau d\tau'  \, \frac{\Sigma_y(\tau,\tau')  \,\Sigma_z(\tau,\tau')}{\Pi(\tau-\tau')} \ .
\end{eqnarray}
The Majorana Green function in \eqref{SetaQMMM} is
\begin{align}\label{eq:Ginv}
 G^{-1} &=  \left(
 \begin{array}{ccc}
- \delta(\tau-\tau') \partial_{\tau'} & i B \delta(\tau-\tau') & 0 \\
 -i B \delta(\tau-\tau') & -\delta(\tau-\tau') \partial_{\tau'} - \Sigma_y(\tau,\tau')& 0 \\
 0 & 0 & -\delta(\tau-\tau') \partial_{\tau'} - \Sigma_z(\tau,\tau')
 \end{array}
 \right)\ .
 \end{align}

The function $\Pi(\tau-\tau')$ is positive and non-zero. The standard form reads 
\begin{equation}
\Pi(\tau-\tau') = \frac{g \pi T^2}{\sin^2(\pi T (\tau-\tau'))}\ .
\end{equation}
The short-time divergence is to be cut off at $|\tau-\tau'| < 1/\Lambda$, leading to the maximal
value of order $g \Lambda^2$. 

The redecoupled action \eqref{SetaQMMM} is again quadratic in the Majorana Grassmann variables
$\eta_\alpha$, which allows us to integrate them out and to obtain an effective action of
the $\Sigma$-fields:
\begin{align} \label{SQ}
 i S[\Sigma_\alpha]= \frac 12  \text{Tr} \log{\left(G^{-1}\right)} 
 - \frac12\int d\tau  d\tau'\,   \frac{\Sigma_y(\tau,\tau')\,\Sigma_z(\tau,\tau')}{\Pi(\tau-\tau')} \ .
 \end{align}

\subsection{Saddle point}

We can now identify the saddle point and fluctuations of the effective $\Sigma$-action.
A saddle-point solution $\Sigma^{sp}_{\alpha}$ is found by expanding
$\Sigma_\alpha=\Sigma^{sp}_{\alpha}+\delta \Sigma_\alpha$ ($\alpha = y,z$) around the saddle point
and taking the linear order in 
$\delta \Sigma_\alpha$. We obtain
\begin{align}
\Sigma^{sp}_{y}(\tau-\tau')&=\Pi(\tau-\tau') G^{sp}_{zz}(\tau-\tau') \,,\\
\Sigma^{sp}_{z}(\tau-\tau')&=\Pi(\tau-\tau') G^{sp}_{yy}(\tau-\tau') \,.
\end{align}
These equations look similar to Eqs.~(\ref{Sigma1y}) and (\ref{Sigma1z}) obtained in the
first order of the diagrammatic expansion (see Fig.~\ref{Sigma1Order}). However, here
in contrast to the first-order calculation, the fermionic lines should be broadened
self-consistently. This means that the Green functions should be calculated using the Dyson
equation 
$[G^{sp}]^{-1} =  [G^{0}]^{-1} - \hat \Sigma^{sp}$. 

Straightforward analysis shows that the self-consistency does not change the result considerably,
and we again obtain the results given by Eqs.~(\ref{Sigma1yResult}) and (\ref{Sigma1zResult}).
However, we are still to investigate fluctuations around the saddle-point solution. 

\subsection{Role of the fluctuations}

\begin{figure}
\includegraphics[scale=0.3]{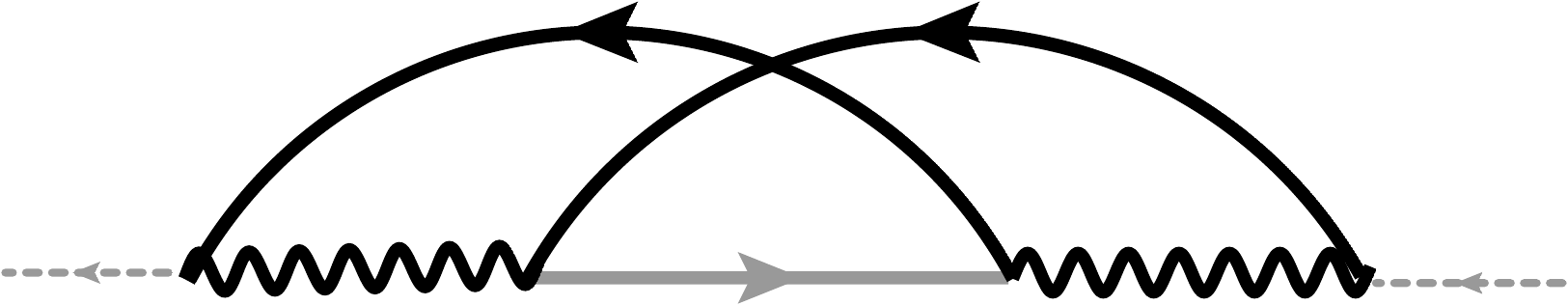}
\includegraphics[scale=0.3]{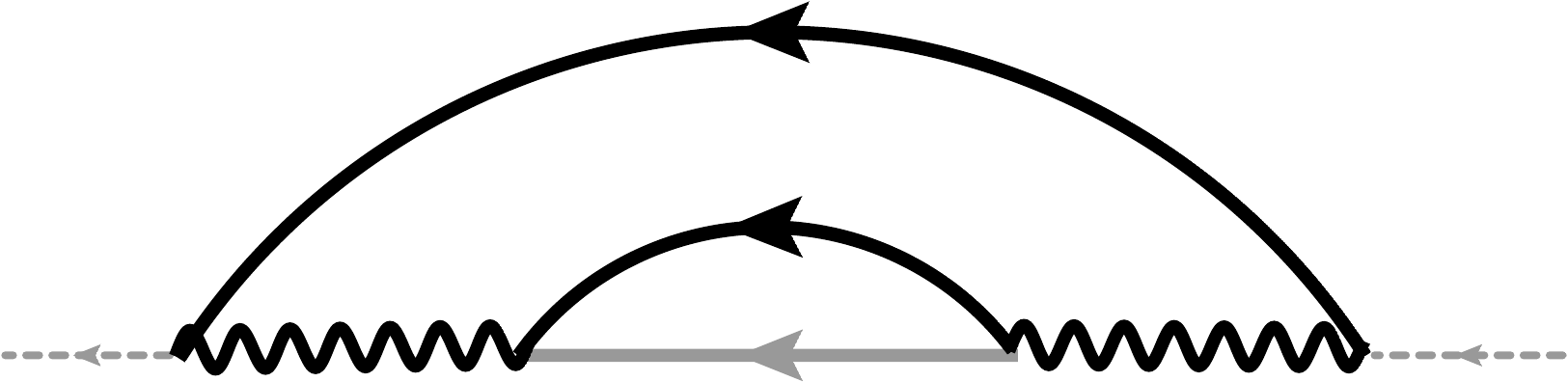}
\caption{Second-order diagrams for the self-energy that correspond to fluctuations of the
self-energy. The colors indicate the two flavors of Majorana fermions ($y$ and $z$). These are the
first two diagrams of Fig.~\ref{Sigma2Order}.}
\label{Sigma2OrderS}
\end{figure}

Reanalyzing the diagrams shown in Fig.~\ref{Sigma2Order} we understand that the third diagram is actually taken into account in the 
saddle-point calculation presented above. The first two diagrams of Fig.~\ref{Sigma2Order}
correspond to fluctuations of the self-energy. This becomes evident if we redraw these diagrams as
shown in Fig.~\ref{Sigma2OrderS}.
Thus these are the fluctuations around the saddle point which are responsible for the divergence of
the longitudinal self-energy $\Sigma_z$. The first-order diagrammatic expansion corresponds to the
saddle-point approximation and is insufficient for the longitudinal spin component.

\section{Generalized Wick theorem}
\label{sec:Wick}

The results obtained above allow us to formulate a prescription for efficient calculation of spin
correlation functions of arbitrary order. This prescription essentially reduces to the suggestion to
avoid $\eta_z$ in all calculations.
More specifically, in order to avoid divergences and the need to account for high-order
contributions, one could use the following approach: for calculation of a spin correlation
function replace $S_x$ with $\eta_x$, $S_y$ with $\eta_y$, and $S_z$ with $\eta_x \eta _y$.


Indeed, consider the effective action  with sources for the Majorana fermions:
\begin{equation}
{\cal S} = \frac{1}{2} \vec\eta^T ([G^{sp}]^{-1} - \delta \Sigma) \vec \eta - \frac{1}{2} \frac{\delta\Sigma_z \delta \Sigma_y}{\Pi} + \vec I \vec \eta\ .
\end{equation}
Here $G^{sp}$ is the Green function at the saddle point, which can be obtained from (\ref{eq:Ginv})
by replacing $\Sigma$ by $\Sigma^{sp}$.
In this expression and below in this section we use sloppy notations implying proper time
integrations. Integrating over the Majorana fermions, we find
$$
{\cal S} = \frac{1}{2} \TrLog \left([G^{sp}]^{-1}-\delta\Sigma\right)-
\frac{1}{2}\frac{\delta\Sigma_z \delta \Sigma_y}{\Pi} 
+ \frac{1}{2} \vec I^T \left([G^{sp}]^{-1}-\delta\Sigma\right)^{-1} \vec I\ .
$$
Since we are going to use only $\eta_x$ and $\eta_y$, only the sources $I_x$ and
$I_y$ will be relevant. In turn, this means that only the $x,y$ components of
$\left([G^{sp}]^{-1}-\delta\Sigma\right)$ will appear in the pre-exponential and will be averaged.
These include only $\delta\Sigma_y$, but not $\delta\Sigma_z$.  

Expansion of the $\TrLog$ term to the second order gives
$$
\delta {\cal S} = \frac{1}{2} 
\left(
\begin{array}{cc}
\delta\Sigma_y & \delta\Sigma_z
\end{array}
\right)
\left(\begin{array}{cc}
G^{sp}_y G^{sp}_y    & \Pi^{-1} \\
\Pi^{-1} & G^{sp}_z G^{sp}_z
\end{array}\right)
\left(
\begin{array}{c}
\delta\Sigma_y \\ \delta\Sigma_z
\end{array}
\right)
$$
In Ref.~\onlinecite{Schad2015} we have shown that the kernel $G^{sp}_z G^{sp}_z$ is small in the case $B=0$.
This observation applies here as well since $B$ does not enter $G^{sp}_z$.
This means that the $\langle\delta\Sigma_y \delta\Sigma_y\rangle$ correlator is small. Since only
$\delta\Sigma_y$ is present in the pre-exponential, we do not have to take into account the
fluctuations. This proves the Wick theorem for the saddle-point Green functions. 

As an example, we employ the prescription to avoid $\eta_z$ and the generalized Wick theorem introduced 
above to compute the longitudinal correlation function $2\langle \left\{S_z(t_1),S_z(t_2)\right\}_+\rangle$. 
Details of the calculation are given in Appendix~\ref{sec:C2zz}. The result reads
\begin{equation}
2\langle \left\{S_z(t_1),S_z(t_2)\right\}_+\rangle
= 4\av{S_z}^2 \cdot  2\pi \delta(\omega) +  \left(1-
4\av{S_z}^2\right) \frac{ 2\Gamma_1 }{\omega^2+\Gamma_1^2} \ .
 \label{CzzBR}
\end{equation}
This is indeed the expected Bloch-Redfield result (cf. the discussion in Sec.~\ref{sec:model}), which is consistent with 
Eq.~(\ref{eq:Ganticipated}). 

\section{Conclusions}

We conclude that the problem noticed first in Ref.~\onlinecite{Shnirman03} is apparently solved by
taking into account fluctuations of the self-energy. In other words, calculations beyond
the saddle-point approximation of Fig.~\ref{Sigma1Order} are needed. For the transverse components
the saddle-point approximation is sufficient. Furthermore, efficient calculation is achieved if one
uses only transverse Majorana operators. 

\section{Acknowledgements}

This research was funded by the German Science Foundation (DFG)
through Grant No. SH 81/2-1 (PS and AS), by the German-Israeli Foundation (GIF) (AS),
and by RSF under grant No.~14-12-00898 (dissipative dynamics; YM).

\appendix
\section{Majorana Representation for Spin-1/2 Operators}
\label{sec:majrep}

The following Majorana representation of the spin-$1/2$ operators was introduced by
Martin~\cite{Martin} in 1959:
\begin{equation}
\label{Majrep}
\hat{S}^\alpha = -\frac{i}{2} \epsilon_{\alpha\beta\gamma} \hat\eta_\beta \hat\eta_\gamma, 
\qquad 
\hat{S}^x= -i\hat\eta_y \hat\eta_z ,
\quad 
\hat{S}^y= -i\hat\eta_z \hat\eta_x ,
\quad 
\hat{S}^z= -i\hat\eta_x \hat\eta_y.
\end{equation}
The Majorana operators are Hermitian, $\hat\eta_\alpha^\dagger=\hat\eta_\alpha$,  and obey the Clifford algebra
$\{\hat\eta_\alpha, \hat\eta_\beta\}=\delta_{\alpha\beta}$, $\hat \eta_\alpha^2=1/2$.
It is easy to check that the representation (\ref{Majrep}) perfectly reproduces the SU$(2)$ algebra
of the spin-1/2 operators $\hat{S}^\alpha$:
\begin{equation}
\left[\hat{S}^\alpha,\hat{S}^\beta\right]=i \epsilon_{\alpha\beta\gamma} \hat{S}^\gamma \quad , \quad \hat {\bf S}^2 = 3/4\ .
\end{equation}
The representation (\ref{Majrep}) can be realized in various Hilbert spaces. The minimal Hilbert
space is 4-dimensional and corresponds to two complex (Dirac) fermions, which we call $\hat c$ and
$\hat d$. In this case one formally breaks the isotropy between the axes by choosing, e.g., $\hat
\eta_x = (\hat c+\hat c^\dag)/\sqrt{2}$,  
$\hat \eta_y = (\hat c-\hat c^\dag)/\sqrt{2}i$, and $\hat \eta_z = (\hat d+\hat
d^\dag)/\sqrt{2}$. Such a choice is termed ``drone"-fermion representation and was used, e.g.,
in Ref.~\onlinecite{Spencer68}. Another, more symmetric option is to 
introduce an 8-dimensional Hilbert space corresponding to three complex fermions, $\hat c_\alpha$, where $\alpha={x,y,z}$.
In this case $\hat \eta_\alpha = (\hat c_\alpha+\hat c_\alpha^\dag)/\sqrt{2}$. As discussed in Ref.~\onlinecite{Schad2015}
the choice of the Hilbert space is irrelevant.

To obtain spin-spin correlation functions using (\ref{Majrep}) one would have to calculate 
4-fermion correlators or, more specifically, fermionic loops. In 2003 an observation was 
made~\cite{Mao03,Shnirman03}, which allowed one to calculate just the (single-particle) Green
functions of the Majorana fermions. Indeed, let us rewrite Eqs.~(\ref{Majrep}):
\begin{equation}
\label{Majrep2}
\hat{S}^\alpha = \hat{\Theta} \hat{\eta}_\alpha, \qquad
\hat{\Theta}=- 2i\hat{\eta}_x\hat{\eta}_y\hat{\eta}_z, \qquad
\hat{\Theta}^2=1/2.
\end{equation}
Apparently, the operator $\hat \Theta$ commutes with
all three Majorana operators $\eta_\alpha$. Hence, it also commutes with the spin operators, and
thus with any physical Hamiltonian; thus the corresponding Heisenberg operator is
time-independent. The average product of a pair of spin operators can
now be represented as
\begin{equation}\label{eq:Trick}
\av{\hat{S}^\alpha(t_1)\hat{S}^\beta(t_2)}
= \av{\hat{\Theta}\hat{\eta}_\alpha(t_1) \hat{\Theta}\hat{\eta}_\beta(t_2)} 
= \frac 12 \av{\hat{\eta}_\alpha(t_1)\hat{\eta}_\beta(t_2)}.
\end{equation}
Thus the average of two spin operators is expressed in terms of the
average of only {\it two} Majorana fermions instead of four.
Implications of this relation for time-ordered correlation functions 
in the Matsubara or Keldysh formalism are discussed in detail in Ref.~\onlinecite{Schad2015}.


\section{Fluctuations in the Keldysh technique}

For the Majorana Green function, the corresponding self-energy, and the bath correlator we use the
following Keldysh structure,
\begin{gather}
 \hat{G}_\alpha(\epsilon)=\begin{pmatrix} G^K_\alpha(\epsilon) & G^R_\alpha(\epsilon) \\ G^A_\alpha(\epsilon) & 0 \end{pmatrix}, \quad \hat{\Sigma}_\alpha(\epsilon)=\begin{pmatrix} 0 & \Sigma^A_\alpha(\epsilon) \\ \Sigma^R_\alpha(\epsilon) & \Sigma^K_\alpha(\epsilon) \end{pmatrix} , \quad
 \hat{\Pi}(\omega)=\begin{pmatrix} \Pi^K(\omega) & \Pi^R(\omega) \\ \Pi^A(\omega) & 0 \end{pmatrix} \ .
\end{gather}
The free Majorana Green functions are
\begin{gather}
G^{0,R}_{yy}(\epsilon)= \frac{1}{2}\left[ \frac{1}{\epsilon-B+i0} + \frac{1}{\epsilon+B+i0} \right]\,,
\quad  
G^{0,R}_{zz}(\epsilon)=  \frac{1}{\epsilon +i0}\,,\quad
G^{K}_{\alpha\alpha}(\epsilon) = \tanh\left(\frac{\beta \epsilon}{2}\right) \left[
G^{R}_{\alpha\alpha}(\epsilon) - G^{A}_{\alpha\alpha}(\epsilon) \right]\,,
\end{gather}
and the spectral representation of the bath correlator $\hat\Pi^{ab}(t,t')=  \av{\mathcal{T}_K
\check X^{a}(t) \check X^{b}(t')}$ in the Keldysh technique is given by
\begin{align}
 \Pi^{R/A}(\omega) = \int_{-\Lambda}^\Lambda \frac{dx}{i \pi} \frac{g x}{\omega +x\pm i0},\qquad
\Pi^K(\omega)
 = \coth\left(\frac{\omega}{2 T}\right) \left( \Pi^R(\omega) -\Pi^A(\omega) \right) \,.
\label{PiKeldysh}
\end{align}

In order to write the action in a Keldysh form, we introduce the vectors $\check
X^T=(X^{cl},X^q)$ and $\check \eta_\alpha^T=(\eta_\alpha^{cl},\eta_\alpha^q)$ (we use the convention
$X^{cl} = (X^u + X^d)/\sqrt{2}$ and $X^{q} = (X^u - X^d)/\sqrt{2}$) and the matrices
$\gamma^{cl}=\mathbb{1}$, $\gamma^{q}=\tau_1 \equiv \tau_x$ in the Keldysh space: 
\begin{align}
 iS_{int} &= - \int_C dt \,  X(t) \eta_y(t) \eta_z(t) = -\frac{1}{\sqrt{2}} \int_{-\infty}^\infty dt \ \left( X^{cl}(t) [\check\eta^T_y(t) \gamma^{q} \check\eta_z(t)] + X^{q}(t) [\check\eta^T_y(t) \gamma^{cl} \check\eta_z(t)] \right) \\
 &= -\frac{1}{\sqrt{2}} \sum_{a=q,cl} \int_{-\infty}^\infty dt \, (\tau_1 \check X(t))^a
[\check\eta^T_y(t) \gamma^{a} \check\eta_z(t)]
\,.
\end{align}

The contribution of the first diagram of Fig. \ref{Sigma2OrderS} to the retarded self-energy
$\Sigma^R=\Sigma^{q,cl}$ reads 
\begin{align}
 \frac{1}{4} \sum_{a,b,c,d=q,cl} \int_{-\infty}^\infty \frac{d\omega_1 d\omega_2}{(2\pi)^2} \,
\left( \gamma^a \hat G_y(\omega_2) \gamma^b \hat{G}_z(\omega_1+\omega_2)\gamma^c \hat
G_y(\omega_1+\epsilon) \gamma^d \right)^{q,cl} \left(\tau_1 \hat \Pi(\omega_2-\epsilon) \tau_1
\right)^{ca}  \left(\tau_1 \hat \Pi(\omega_1) \tau_1 \right)^{db}  \,,
\end{align}
while the second diagram contributes
\begin{align}
 -\frac{1}{4} \sum_{a,b,c,d=q,cl} \int_{-\infty}^\infty \frac{d\omega_1 d\omega_2}{(2\pi)^2} \,
\trace\{ \hat G_y(\omega_2) \gamma^b  \hat{G}_z(\omega_1+\omega_2) \gamma^c \} \left(\gamma^a \hat
G_y(\omega_1+\epsilon) \gamma^d \right)^{q,cl} \left(\tau_1 \hat \Pi(\omega_1) \tau_1 \right)^{ca} 
\left(\tau_1 \hat \Pi(\omega_1) \tau_1 \right)^{db}  \,.
\end{align}

Both diagrams were evaluated using the spectral representation \eqref{PiKeldysh}. We found (with
the help of Mathematica) that the result agrees with that obtained in the Matsubara technique and
therefore confirms that the two diagrams in Fig.~\ref{Sigma2OrderS} are responsible for the
divergent term in the expansion (\ref{eq:SigmaExpanded}).


\section{Longitudinal spin correlations}
\label{sec:C2zz}

In this appendix we demonstrate how the generalized Wick theorem can be utilized in practice. To this end, we compute the longitudinal spin-spin correlation function  $C^{(2)}_{zz}(t,t')=2\av{{\mc T}_K S^{cl}_z(t) S^{cl}_z(t')}$ employing the Keldysh technique.  We avoid the longitudinal Majorana fermion $\eta_z$ and use instead the relation $S_z=-i\eta_x\eta_y$.
The correlator is, thus, given by  
\begin{align}
 C^{(2)}_{zz}(t,t')&= -\av{{\mc T}_K [\check\eta_x(t)\gamma^{cl}\check\eta_y(t)]
[\check\eta_x(t')\gamma^{cl}\check\eta_y(t')]} \,.
 \label{C2zNm}
\end{align}
The generalized Wick theorem for saddle-point Green functions means that vertex corrections to the
$z$-spin vertices are absent. Thus computing \eqref{C2zNm} we obtain three contributions, which
correspond to the diagrams in Fig.~\ref{fig:C2z}:
\begin{figure}
\centering 
a)\includegraphics[scale=.5]{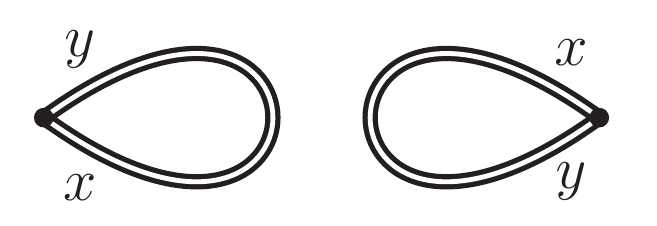} \qquad\qquad
b)\includegraphics[scale=.5]{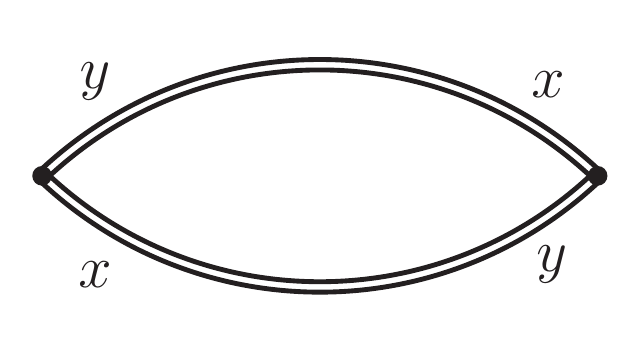} \qquad\qquad
c)\includegraphics[scale=.5]{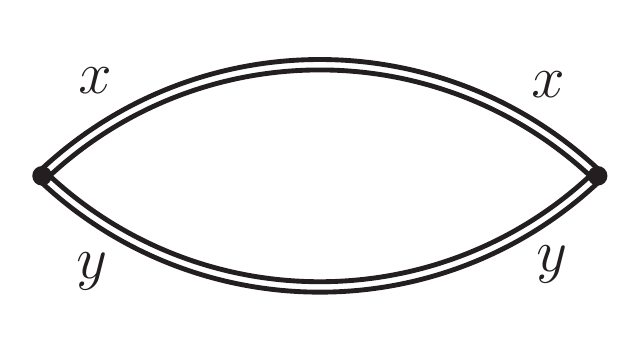}   
\caption{Three contributions to the longitudinal spin correlation function. Double lines: fermionic saddle-point Green functions.}
\label{fig:C2z}
\end{figure}
%
\begin{multline}
 C^{(2)}_{zz}(\omega) = 2\pi \delta(\omega) \left( \int\frac{d\Omega}{2\pi} \,
G^{sp,K}_{xy}(\Omega) \right)^2 
 - \int \frac{d\Omega}{2\pi} \trace\left[\hat G^{sp}_{xy}(\Omega) \hat
G^{sp}_{xy}(\Omega-\omega)\right] 
 +   \int \frac{d\Omega}{2\pi} \trace\left[\hat G^{sp}_{xx}(\Omega) \hat G^{sp}_{yy}(\Omega-\omega
)\right] \,.
 \label{C2zNmf}
\end{multline}
Here the traces are taken over the Keldysh indices. 
In order to perform the calculation we need the saddle-point Green functions of Majorana fields
$\eta_x$ and $\eta_y$. 
It is sufficient to calculate the retarded components of these Green functions, and we obtain 
\begin{gather}
 G^{sp,R}_{xx}(\epsilon)=\frac{\epsilon-\Sigma_y^{sp,R}(\epsilon)}{\epsilon
\left(\epsilon-\Sigma_y^{sp,R}(\epsilon) \right) -B^2} \,, \qquad 
 G^{sp,R}_{yy}(\epsilon)=\frac{\epsilon}{\epsilon \left(\epsilon-\Sigma_y^{sp,R} (\epsilon) \right) -B^2}
\,, \notag\\[1ex]
 G^{sp,R}_{xy}(\epsilon)=\frac{-iB }{\epsilon \left(\epsilon-\Sigma_y^{sp,R}(\epsilon) \right) -B^2} \,,
\qquad\qquad G^{sp,R}_{yx}(\epsilon)=-G^{sp,R}_{xy}(\epsilon) \,.
 \label{GxyspSB}
\end{gather}
Here the retarded self-energy $\Sigma_y^{sp,R}(\epsilon)$ is (approximately) given by the analytic continuation of (\ref{Sigma1yResult}). It is sufficient to replace $\Sigma_y^{sp,R}(\epsilon)$ by its value in the vicinity of the poles $\epsilon \approx \pm B$, i.e., 
$\Sigma_y^{sp,R}(\epsilon) \rightarrow -i\Gamma_1$ (cf. (\ref{Sigma1yRetarded})).

We can now calculate the longitudinal spin correlator $C^{(2)}_{zz}(\omega)$. We consider the limit of high magnetic field 
$B\gg \Gamma_1$.
The first term in \eqref{C2zNmf}, corresponding to Fig.~\ref{fig:C2z}a, yields a factor of $A =
\tanh^2\frac{B}{2T} = 4\av{S_z}^2$. The second and third terms in \eqref{C2zNmf}, i.e.,
Fig.~\ref{fig:C2z}b,c, both give rise to a Lorentzian of width $\Gamma_1$. Thus we obtain
\begin{equation}
 C^{(2)}_{zz}(\omega)= A \cdot 2\pi \delta(\omega)  +
 \frac{\left(1- A\right) 2\Gamma_1}{\omega^2+\Gamma_1^2} \ ,
 \label{Czznm}
\end{equation}
which indeed coincides with the Bloch-Redfield result discussed in Sec.~\ref{sec:model}. 

\bibliography{annals}

\begin{thebibliography}{26}
\expandafter\ifx\csname natexlab\endcsname\relax\def\natexlab#1{#1}\fi
\expandafter\ifx\csname bibnamefont\endcsname\relax
  \def\bibnamefont#1{#1}\fi
\expandafter\ifx\csname bibfnamefont\endcsname\relax
  \def\bibfnamefont#1{#1}\fi
\expandafter\ifx\csname citenamefont\endcsname\relax
  \def\citenamefont#1{#1}\fi
\expandafter\ifx\csname url\endcsname\relax
  \def\url#1{\texttt{#1}}\fi
\expandafter\ifx\csname urlprefix\endcsname\relax\def\urlprefix{URL }\fi
\providecommand{\bibinfo}[2]{#2}
\providecommand{\eprint}[2][]{\url{#2}}

\bibitem[{\citenamefont{Altland and Simons}(2010)}]{altland}
\bibinfo{author}{\bibfnamefont{A.}~\bibnamefont{Altland}} \bibnamefont{and}
  \bibinfo{author}{\bibfnamefont{B.}~\bibnamefont{Simons}},
  \emph{\bibinfo{title}{Condensed Matter Field Theory}}
  (\bibinfo{publisher}{Cambridge University Press}, \bibinfo{year}{2010}),
  \bibinfo{edition}{2nd} ed.

\bibitem[{\citenamefont{Tsvelik}(1996)}]{Tsvelik}
\bibinfo{author}{\bibfnamefont{A.~M.} \bibnamefont{Tsvelik}},
  \emph{\bibinfo{title}{Quantum field theory in condensed matter physics}}
  (\bibinfo{publisher}{Cambridge University Press},
  \bibinfo{address}{Cambridge}, \bibinfo{year}{1996}).

\bibitem[{\citenamefont{Jordan and Wigner}(1928)}]{JordanWigner}
\bibinfo{author}{\bibfnamefont{P.}~\bibnamefont{Jordan}} \bibnamefont{and}
  \bibinfo{author}{\bibfnamefont{E.}~\bibnamefont{Wigner}},
  \bibinfo{journal}{Zeitschrift f{\"u}r Physik} \textbf{\bibinfo{volume}{47}},
  \bibinfo{pages}{631} (\bibinfo{year}{1928}).

\bibitem[{\citenamefont{Holstein and Primakoff}(1940)}]{Holstein}
\bibinfo{author}{\bibfnamefont{T.}~\bibnamefont{Holstein}} \bibnamefont{and}
  \bibinfo{author}{\bibfnamefont{H.}~\bibnamefont{Primakoff}},
  \bibinfo{journal}{Phys. Rev.} \textbf{\bibinfo{volume}{58}},
  \bibinfo{pages}{1098} (\bibinfo{year}{1940}).

\bibitem[{\citenamefont{Martin}(1959)}]{Martin}
\bibinfo{author}{\bibfnamefont{J.~L.} \bibnamefont{Martin}},
  \bibinfo{journal}{Proc. R. Soc. London A} \textbf{\bibinfo{volume}{251}},
  \bibinfo{pages}{536} (\bibinfo{year}{1959}).

\bibitem[{\citenamefont{Abrikosov}(1965)}]{abrikosov}
\bibinfo{author}{\bibfnamefont{A.~A.} \bibnamefont{Abrikosov}},
  \bibinfo{journal}{Physics} \textbf{\bibinfo{volume}{2}}, \bibinfo{pages}{5}
  (\bibinfo{year}{1965}).

\bibitem[{\citenamefont{Schwinger}(1965)}]{schwinger}
\bibinfo{author}{\bibfnamefont{J.}~\bibnamefont{Schwinger}}, in
  \emph{\bibinfo{booktitle}{Quantum Theory of Angular Momentum}}, edited by
  \bibinfo{editor}{\bibfnamefont{L.}~\bibnamefont{Biederharn}}
  \bibnamefont{and} \bibinfo{editor}{\bibfnamefont{H.}~\bibnamefont{van Dam}}
  (\bibinfo{publisher}{Academic Press}, \bibinfo{address}{New York},
  \bibinfo{year}{1965}), p. \bibinfo{pages}{229}.

\bibitem[{\citenamefont{Arovas and Auerbach}(1988)}]{arovas}
\bibinfo{author}{\bibfnamefont{D.~P.} \bibnamefont{Arovas}} \bibnamefont{and}
  \bibinfo{author}{\bibfnamefont{A.}~\bibnamefont{Auerbach}},
  \bibinfo{journal}{Phys. Rev. B} \textbf{\bibinfo{volume}{38}},
  \bibinfo{pages}{316} (\bibinfo{year}{1988}).

\bibitem[{\citenamefont{Read and Sachdev}(1991)}]{read}
\bibinfo{author}{\bibfnamefont{N.}~\bibnamefont{Read}} \bibnamefont{and}
  \bibinfo{author}{\bibfnamefont{S.}~\bibnamefont{Sachdev}},
  \bibinfo{journal}{Phys. Rev. Lett.} \textbf{\bibinfo{volume}{66}},
  \bibinfo{pages}{1773} (\bibinfo{year}{1991}).

\bibitem[{\citenamefont{Wang}(2010)}]{wang}
\bibinfo{author}{\bibfnamefont{F.}~\bibnamefont{Wang}}, \bibinfo{journal}{Phys.
  Rev. B} \textbf{\bibinfo{volume}{82}}, \bibinfo{pages}{024419}
  (\bibinfo{year}{2010}).

\bibitem[{\citenamefont{Affleck et~al.}(1988)\citenamefont{Affleck, Zou, Hsu,
  and Anderson}}]{affleck}
\bibinfo{author}{\bibfnamefont{I.}~\bibnamefont{Affleck}},
  \bibinfo{author}{\bibfnamefont{Z.}~\bibnamefont{Zou}},
  \bibinfo{author}{\bibfnamefont{T.}~\bibnamefont{Hsu}}, \bibnamefont{and}
  \bibinfo{author}{\bibfnamefont{P.~W.} \bibnamefont{Anderson}},
  \bibinfo{journal}{Phys. Rev. B} \textbf{\bibinfo{volume}{38}},
  \bibinfo{pages}{745} (\bibinfo{year}{1988}).

\bibitem[{\citenamefont{Marston and Affleck}(1989)}]{marston}
\bibinfo{author}{\bibfnamefont{J.~B.} \bibnamefont{Marston}} \bibnamefont{and}
  \bibinfo{author}{\bibfnamefont{I.}~\bibnamefont{Affleck}},
  \bibinfo{journal}{Phys. Rev. B} \textbf{\bibinfo{volume}{39}},
  \bibinfo{pages}{11538} (\bibinfo{year}{1989}).

\bibitem[{\citenamefont{Andrei and Coleman}(1989)}]{andrei}
\bibinfo{author}{\bibfnamefont{N.}~\bibnamefont{Andrei}} \bibnamefont{and}
  \bibinfo{author}{\bibfnamefont{P.}~\bibnamefont{Coleman}},
  \bibinfo{journal}{Phys. Rev. Lett.} \textbf{\bibinfo{volume}{62}},
  \bibinfo{pages}{595} (\bibinfo{year}{1989}).

\bibitem[{\citenamefont{Dagotto et~al.}(1988)\citenamefont{Dagotto, Fradkin,
  and Moreo}}]{dagotto}
\bibinfo{author}{\bibfnamefont{E.}~\bibnamefont{Dagotto}},
  \bibinfo{author}{\bibfnamefont{E.}~\bibnamefont{Fradkin}}, \bibnamefont{and}
  \bibinfo{author}{\bibfnamefont{A.}~\bibnamefont{Moreo}},
  \bibinfo{journal}{Phys. Rev. B} \textbf{\bibinfo{volume}{38}},
  \bibinfo{pages}{2926} (\bibinfo{year}{1988}).

\bibitem[{\citenamefont{Wen}(1991)}]{wen}
\bibinfo{author}{\bibfnamefont{X.~G.} \bibnamefont{Wen}},
  \bibinfo{journal}{Phys. Rev. B} \textbf{\bibinfo{volume}{44}},
  \bibinfo{pages}{2664} (\bibinfo{year}{1991}).

\bibitem[{\citenamefont{Burnell and Nayak}(2011)}]{nayak}
\bibinfo{author}{\bibfnamefont{F.~J.} \bibnamefont{Burnell}} \bibnamefont{and}
  \bibinfo{author}{\bibfnamefont{C.}~\bibnamefont{Nayak}},
  \bibinfo{journal}{Phys. Rev. B} \textbf{\bibinfo{volume}{84}},
  \bibinfo{pages}{125125} (\bibinfo{year}{2011}).

\bibitem[{\citenamefont{Spencer}(1968)}]{Spencer68}
\bibinfo{author}{\bibfnamefont{H.~J.} \bibnamefont{Spencer}},
  \bibinfo{journal}{Phys. Rev.} \textbf{\bibinfo{volume}{171}},
  \bibinfo{pages}{515} (\bibinfo{year}{1968}).

\bibitem[{\citenamefont{Schad et~al.}(2015)\citenamefont{Schad, Makhlin,
  Narozhny, Sch{\"o}n, and Shnirman}}]{Schad2015}
\bibinfo{author}{\bibfnamefont{P.}~\bibnamefont{Schad}},
  \bibinfo{author}{\bibfnamefont{Y.}~\bibnamefont{Makhlin}},
  \bibinfo{author}{\bibfnamefont{B.}~\bibnamefont{Narozhny}},
  \bibinfo{author}{\bibfnamefont{G.}~\bibnamefont{Sch{\"o}n}},
  \bibnamefont{and} \bibinfo{author}{\bibfnamefont{A.}~\bibnamefont{Shnirman}},
  \bibinfo{journal}{Annals of Physics} \textbf{\bibinfo{volume}{361}},
  \bibinfo{pages}{401 } (\bibinfo{year}{2015}).

\bibitem[{\citenamefont{Mao et~al.}(2003)\citenamefont{Mao, Coleman, Hooley,
  and Langreth}}]{Mao03}
\bibinfo{author}{\bibfnamefont{W.}~\bibnamefont{Mao}},
  \bibinfo{author}{\bibfnamefont{P.}~\bibnamefont{Coleman}},
  \bibinfo{author}{\bibfnamefont{C.}~\bibnamefont{Hooley}}, \bibnamefont{and}
  \bibinfo{author}{\bibfnamefont{D.}~\bibnamefont{Langreth}},
  \bibinfo{journal}{Phys. Rev. Lett.} \textbf{\bibinfo{volume}{91}},
  \bibinfo{pages}{207203} (\bibinfo{year}{2003}).

\bibitem[{\citenamefont{Shnirman and Makhlin}(2003)}]{Shnirman03}
\bibinfo{author}{\bibfnamefont{A.}~\bibnamefont{Shnirman}} \bibnamefont{and}
  \bibinfo{author}{\bibfnamefont{Y.}~\bibnamefont{Makhlin}},
  \bibinfo{journal}{Phys. Rev. Lett.} \textbf{\bibinfo{volume}{91}},
  \bibinfo{pages}{207204} (\bibinfo{year}{2003}).

\bibitem[{\citenamefont{Leggett et~al.}(1987)\citenamefont{Leggett,
  Chakravarty, Dorsey, Fisher, Garg, and Zwerger}}]{LeggettSB87}
\bibinfo{author}{\bibfnamefont{A.~J.} \bibnamefont{Leggett}},
  \bibinfo{author}{\bibfnamefont{S.}~\bibnamefont{Chakravarty}},
  \bibinfo{author}{\bibfnamefont{A.~T.} \bibnamefont{Dorsey}},
  \bibinfo{author}{\bibfnamefont{M.~P.~A.} \bibnamefont{Fisher}},
  \bibinfo{author}{\bibfnamefont{A.}~\bibnamefont{Garg}}, \bibnamefont{and}
  \bibinfo{author}{\bibfnamefont{W.}~\bibnamefont{Zwerger}},
  \bibinfo{journal}{Rev. Mod. Phys.} \textbf{\bibinfo{volume}{59}},
  \bibinfo{pages}{1} (\bibinfo{year}{1987}).

\bibitem[{\citenamefont{Bloch}(1957)}]{Bloch57}
\bibinfo{author}{\bibfnamefont{F.}~\bibnamefont{Bloch}},
  \bibinfo{journal}{Phys. Rev.} \textbf{\bibinfo{volume}{105}},
  \bibinfo{pages}{1206} (\bibinfo{year}{1957}).

\bibitem[{\citenamefont{Redfield}(1957)}]{Redfield57}
\bibinfo{author}{\bibfnamefont{A.}~\bibnamefont{Redfield}},
  \bibinfo{journal}{IBM Journal of Research and Development}
  \textbf{\bibinfo{volume}{1}}, \bibinfo{pages}{19} (\bibinfo{year}{1957}).

\bibitem[{\citenamefont{Florens et~al.}(2011)\citenamefont{Florens, Freyn,
  Venturelli, and Narayanan}}]{FlorensPRB11}
\bibinfo{author}{\bibfnamefont{S.}~\bibnamefont{Florens}},
  \bibinfo{author}{\bibfnamefont{A.}~\bibnamefont{Freyn}},
  \bibinfo{author}{\bibfnamefont{D.}~\bibnamefont{Venturelli}},
  \bibnamefont{and}
  \bibinfo{author}{\bibfnamefont{R.}~\bibnamefont{Narayanan}},
  \bibinfo{journal}{Phys. Rev. B} \textbf{\bibinfo{volume}{84}},
  \bibinfo{pages}{155110} (\bibinfo{year}{2011}).

\bibitem[{\citenamefont{Vieira}(1981)}]{VieiraPRB81}
\bibinfo{author}{\bibfnamefont{V.~R.} \bibnamefont{Vieira}},
  \bibinfo{journal}{Phys. Rev. B} \textbf{\bibinfo{volume}{23}},
  \bibinfo{pages}{6043} (\bibinfo{year}{1981}).

\bibitem[{\citenamefont{Vieira}(1982)}]{VieiraPhysA82}
\bibinfo{author}{\bibfnamefont{V.~R.} \bibnamefont{Vieira}},
  \bibinfo{journal}{Physica A: Statistical Mechanics and its Applications}
  \textbf{\bibinfo{volume}{115}}, \bibinfo{pages}{58} (\bibinfo{year}{1982}).

\end{thebibliography}

\end{document}